# Persistent chlorophyll maxima in the Eastern Tropical North Pacific

**Amaru Márquez-Artavia[1], Xiomara M. Márquez-Artavia[2], Juan P. Salazar-Ceciliano[2],**

**Aurélien Paulmier[3], Laura Sánchez-Velasco[4] and Emilio Beier[5]**

[1]Laboratorio de Macroecología Marina-CONACyT. Centro de Investigación Científica y de Educación Superior de Ensenada. Unidad La Paz. La Paz, BCS, México.

[2]Universidad Nacional. Departamento de Física. Heredia, Costa Rica.

[3]Laboratoire d'Etudes en Géophysique et Océanographie Spatiales (LEGOS), IRD/CNRS/UPS/CNES, Université de Toulouse, 31401 Toulouse Cedex 9, France

[4]Departamento de Plancton y Ecología Marina. Instituto Politécnico Nacional-Centro Interdisciplinario de Ciencias Marinas. La Paz, BCS, México. C.P. 23096.

[5]Laboratorio de Macroecología Marina. Centro de Investigación Científica y de Educación Superior de Ensenada. Unidad La Paz. La Paz, BCS, México.

**Corresponding author:** Emilio Beier (ebeier@cicese.mx)

**Abstract**

This study aims to describe the response of two persistent chlorophyll-*a* maxima to physical processes affecting the thermocline/nitracline position in the Eastern Tropical North Pacific (ETNP). We focused on Long Rossby Waves given their relevance to the ETNP circulation and its potential as a mechanism introducing nutrients into the euphotic zone. We found the shallower chlorophyll-*a* maximum in oxygenated waters became more intense when denser waters (more nutrients) moved toward the surface. It suggests that isopycnals and nitracline displacements modify the nutrient supply in the euphotic zone, which produces changes in phytoplankton growth. The suboxic and deeper chlorophyll-*a* maximum showed a strong association with the 26 kg m$^{-3}$ isopycnal being only mechanically displaced, and its chlorophyll-*a* content does not seem to covary with irradiance or nutrients. The different responses of the chlorophyll-*a* maxima could be explained if different phytoplankton groups are associated with them. Long Rossby Waves can affect the position of the thermocline/nitracline and isopycnals in an annual cycle, but it seems to be a "*background*" signal modulated by higher frequency processes such as mesoscale eddies and other Rossby waves. The co-occurrence of processes can control the nitracline depth, and hence the input of nutrients into the euphotic zone that can cause sporadic enhancements of the chlorophyll-*a* concentration of one maximum.

1. **Introduction**

Chlorophyll-*a* maxima are relevant features of the pelagic ecosystem, which can originate from several processes. For example, they can form near the nitracline due to an increase in the phytoplankton growth; a scenario in which they coincided with the biomass maximum and occurring typically in stratified waters with mesotrophic or eutrophic conditions (Cullen, 2015). Chlorophyll-*a* maxima can also develop by photoacclimation; phytoplankton increased their intracellular pigment concentration in response to the low-light availability (Cullen, 2015; Mignot et al., 2014). This case is commonly found in oligotrophic regions, and the chlorophyll-*a* maximum only represents changes in the pigment concentration, not in biomass.

In the Oxygen Minimum Zones (OMZs), such as the Eastern Tropical North Pacific (ETNP) and the Arabian Sea, several chlorophyll-*a* maxima can develop in the water column (Garcia-Robledo et al., 2017; Goericke et al., 2000; Lavin et al., 2010; Márquez-Artavia et al., 2019). A chlorophyll-*a* maximum forms near the nutricline in well-oxygenated waters, while another maximum develops below the nutricline, in the lower part of the euphotic zone, and within suboxic waters ($< 20$ µmol$O_2$ kg$^{-1}$). This maximum found in suboxic waters is caused by an increase in the cell number of *Prochlorococcus* ecotypes of the OMZs (Goericke et al., 2000; Lavin et al., 2010). This chlorophyll-*a* maximum is of biogeochemical relevance because it constitutes a source of organic matter and oxygen that is locally consumed (Garcia-Robledo et al., 2017), but little is known about the physical mechanisms that affect and maintain these phytoplankton populations.

Some physical processes affecting phytoplankton populations have been reviewed in the literature, including the effects of mesoscale eddies and the Long Rossby Waves (LRWs) (Killworth et al., 2004; McGillicuddy, 2016; Siegel et al., 1999; Uz et al., 2001). Both processes modify the vertical distribution of properties like nutrients, oxygen, and chlorophyll-*a* (McGillicuddy, 2016; McGillicuddy et al., 2007; Sakamoto et al., 2004), but with different spatial and temporal scales. The effects of LRWs on the distribution of chlorophyll-*a* can be physical or biological (Killworth et al., 2004). Physical effects can be considered as mechanical displacements by horizontal or vertical advection (Belonenko et al., 2018; Killworth et al., 2004), while the biological processes are related to phytoplankton responses to the nutrient supply and light availability. In this regard, Uz et al. (2001) proposed that LRWs can pump nutrients into the euphotic zone, promoting an increase in chlorophyll-*a* concentration.

In the OMZ of the Arabian Sea, it has been documented that the LRWs changed the depth and intensity of the two chlorophyll-*a* maxima, one of which occurs sporadically within the suboxic waters (Ravichandran et al., 2012). In the ETNP, LRWs have been described as an important process for the regional circulation, controlling the position of the thermocline in seasonal scales (Kessler, 2006, 1990). Indeed, the thermocline variability produced by LRWs has maximum amplitudes at 13 ºN (Capotondi et al., 2003), but the effects of these waves on the chlorophyll-*a* maxima have not been assessed yet, despite the extensive review of the chlorophyll-*a* and primary productivity carried by Pennington et al. (2006). In addition, these authors do not consider the formation and variability of the chlorophyll-*a* maximum found in suboxic waters, which is a persistent feature in some regions of the ETNP (Márquez-Artavia et al., 2019)

The aim of this study is to describe the temporal evolution of phytoplankton communities indicated by the chlorophyll-*a* concentration in the OMZ of the ETNP, taking advantage of the high resolution and relatively long record available from a Biogeochemical Argo float (BGC-Argo float). The data set used here allows us to resolve more processes than the monthly climatologies of the ship and satellite-based observations previously used by Pennington et al. (2006). We paid attention to the contribution of LRWs because of their relevance at 13º N and their potential as a physical mechanism introducing nutrients into the euphotic zone. Considering this, we tested the following hypothesis: When LRWs displace isopycnals towards the surface (cyclonic phase) they will increase the nutrient supply into the euphotic zone, causing an increase in phytoplankton growth. The opposite scenario will occur when LRWs deepen the isopycnals (anti-cyclonic phase).

**2. Data and methods**

2.1. Data Sources

We used vertical profiles gathered by a BGC Argo Float (WMO3901531) from 30th November 2016 to 10th March 2019. The float was deployed 110 ºW and 13 ºN and it was programmed to surface at local noon after drifting for five days at 1000 meters. The BGC-Argo float trajectory showed several loops, but the stations were mostly located along the east-west axis with a mean position at 112.16 °W and 13.19 °N (Figure 1). This nearly zonal trajectory of the float allowed us to compare with latitudinal dependent processes such as Rossby waves.

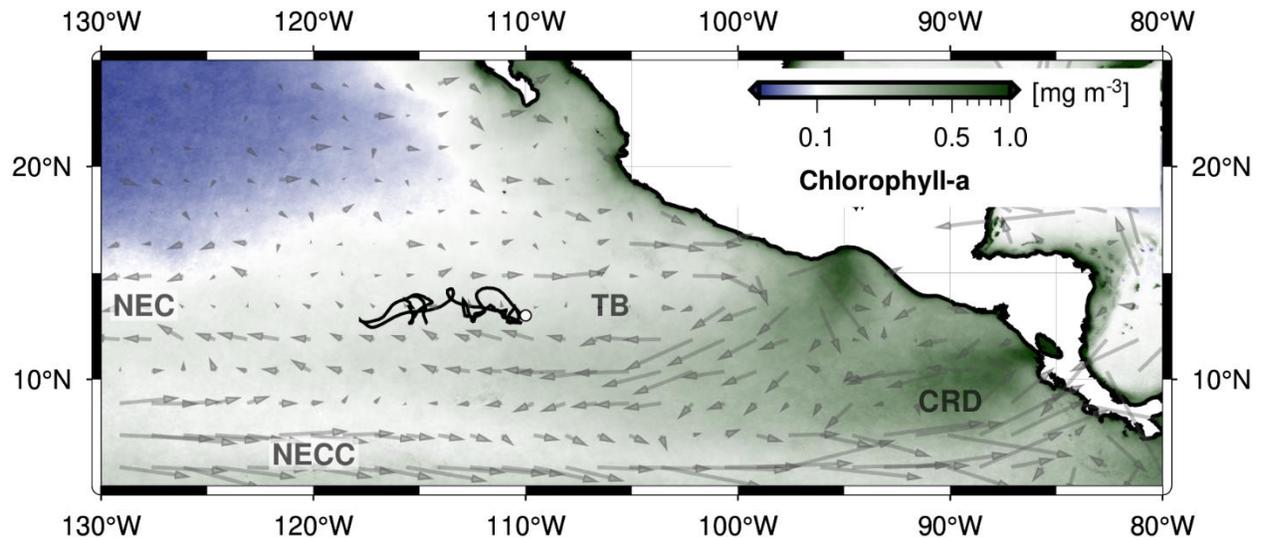

**Figure 1.** Representation of main oceanographic features of the ETNP and the trajectory of the BGC-Argo float. Surface chlorophyll-a measured by satellite radiometers and geostrophic currents from altimetry were averaged using the 2014-2018 period. The white point represents the deployment location of the float and the black line its trajectory. The position of the North Equatorial Counter Current (NECC), Costa Rica Dome (CRD), Tehuantepec Bowl (TB) and the North Equatorial Current are also indicated. Chlorophylll-a data correspond to the version 5 of the ESA-CCI ocean color product (Sathyendranath et al. 2019).

We noted in figure 1, that the float sampled in oceanic waters with a mean surface chlorophyll-*a* slightly above 0.1 mg m$^{-3}$, which corresponds to mesotrophic conditions according to the classification of Antoine et al. (1996). Besides the BGC-Argo float trajectory was between the western flank of an anticyclonic circulation known as the Tehuantepec Bowl (centered at 13 ºN and 105 ºW), and the beginning of the North Equatorial Current, which can be identifiable as a zonal flow to the west of 125 ºW (Figure 1). The float mean speed, estimated from station locations and time was 3.6 ± 2.3 cm s$^{-1}$, similar to the speed at 1000 m  (2.88 ± 1.70 cm s$^{-1}$) reported in the ANDRO dataset (Ollitrault and Rannou, 2013). Thus, the float sampled at the boundaries of the ETNP, and its trajectory was dominated by intermediate-depth currents, just as can be expected of an instrument drifting for five days at 1000 m.

The WMO3901531 BGC-Argo float is equipped with sensors to measure temperature, salinity, oxygen, chlorophyll-*a* fluorescence, light backscattering at an angle of 124º at two wavelengths: 532 and 700 nm. It also measures irradiance in several channels (380, 412, 490 nm and integrated into Photosynthetically Active Radiation spectral band; PAR). Each sensor has its own sampling scheme and resolution. Because of this, we used the synthetic profile in which all variables are reported in the same pressure axis. The detailed procedure to build these synthetic profiles can be found in Bittig et al., (2018) and the data are available from the Coriolis Global Data Assembly Center (ftp://ftp.ifremer.fr/ifremer/argo/etc/argo-synthetic-profile).

We obtained the Sea Surface Level Anomalies (SLA) from the delayed mode multi-mission altimeter gridded fields Level-4 with a resolution of 0.25ºx0.25º from January 1$^{st}$, 2014 to May 13$^{th}$ 2019, and they were downloaded from the Copernicus Marine Environment Monitoring Service (http://marine.copernicus.eu). We used SLA data to analyse if the observed variability in the float time series can be linked to oceanographic phenomena of meso and large scales such as eddies and Rossby waves. We calculated the LRWs phase speed from a longitude-time diagram, by visually defining lines that followed the slanted features and estimating their slope (Barron et al., 2009; Glatt et al., 2011).

In order to identify mesoscale eddies we used the py-eddy-tracker software, which detects eddies based on closed contours of SLA (Mason et al., 2014). Closed contours are computed every 1 cm in the range of -100 to 100 cm. Each closed contour in the interval is compared with a fitted circle with the same area, and it is determined the shape error. Shape error is the sum of the areas that depart from the fitted circle and expressed as a percentage of the total area (see figure 3 in Kurian et al., 2011). Thus, for a given closed contour to be identified as an eddy, it should accomplish the following criteria: (i) the shape error cannot exceed 55%, (ii) the number of pixels within the closed contour should be between 8 and 100, (iii) all SLA values should be above or below from those defined by the closed contour, (iv) there will be only a local maximum or minimum within the closed contour and, (v) the amplitudes should be 1-150 cm. Eddies defined by the closed contours are rather irregular than perfect circles, defining the so-called effective contour, which we used to determine if the BGC-Argo float was inside or outside of a given structure. Finally, we made eddy identification on raw SLA, and we do not screen by any eddy characteristics like the lifespan or size.

2.2 BGC-ARGO float data handling

We used temperature and salinity records from the synthetic profiles to compute Absolute Salinity, Conservative temperature and the potential density anomaly referred to the surface, using the thermodynamic equation of seawater, TEOS-10 (Mcdougall and Barker, 2017). The depth of the 20 °C isotherm (Z20) was selected as a proxy of the thermocline depth to compare with model predictions and following the previous work of Kessler (1990, 2006). We also calculated the pycnocline depth to show that Z20 represents the dynamics of the pycnocline in ETNP near 13 °N. Pycnocline depth was calculated as the maximum gradient of density relative to depth.

We obtained the vertical distribution of nitrate using the temperature, salinity, oxygen, geographical position, and the date of a given profile gathered by the BGC-ARGO float, by using the CANYON algorithm developed by Sauzède et al. (2017). Despite CANYON being a convenient method to obtain nitrate concentration when direct measurements are not available, it gives a biased estimation in the OMZ, probably because it cannot represent the denitrification taking place in the oxygen-deficient waters (see supplementary material). Because of that we only used the CANYON nitrate profiles to compute the nitracline depth, computed as the maximum vertical gradient. The depth of Z20, pycnocline, and nitracline were correlated using Pearson's coefficient.

The annual cycles Z20, pycnocline, nitracline and the chlorophyll-*a* concentration were obtained by least squares fitting of a sinusoidal function such as:

$$y(t) = \underline{y} + A\,cos(\omega t - \phi)$$

where $\underline{y}$ is the mean of the record estimated from the fitting, A and $\phi$ are coefficients denoting the amplitude and the phase of the cosine function, while $\omega$ is the angular frequency ($2\pi/365.25$ days) and *t* is the time. The least-square fitting was done according to Ripa (2002) to get the uncertainties of the coefficients.

Chlorophyll-*a* from fluorescence and light backscattering were processed and quality controlled by the Argo Data Management Team (Schmechtig et al., 2019; Schmechtig et al., 2018a; Schmechtig et al., 2018b). For chlorophyll-*a* fluorescence, it includes negative spike removal, the non-photochemical quenching correction, and the application of a global correction factor, to get the least biased estimation of chlorophyll-*a* as described by Roesler et al. (2017). For the light backscattering, the Argo program estimates the particle backscattering coefficient ($b_{bp}$) at single angles according to Boss & Pegau (2001), and the quality control includes the negative spike removal. Despite the BGC-float measure $b_{bp}$ at 532 and 700 nm, only the latter was used as a proxy of particulate organic carbon or phytoplankton biomass, because this wavelength is less sensitive to absorption of pigments and dissolved materials in the seawater (Boss & Häentjens, 2016). We smoothed the chlorophyll-*a* and bbp700 data by using two consecutive running median filters of five and seven points (Briggs et al., 2011; Rembauville et al., 2017). Considering the mean vertical resolution of the synthetic profiles (0.8 m), it is equivalent to make averages in layers of approximately four and six meters.

Quality control of irradiance of the Argo program only applies a global test to identify values outside the expected range (Poteau et al., 2019). Additionally, we implemented the protocol described by Organelli et al. (2016). After this quality control, only 32% of the profiles measured at PAR were of acceptable quality (good or probably good). We considered this a percentage too low. Hence, we obtained profile of PAR irradiance based on the Lambert-Beer law and estimating the attenuation coefficient (KDPAR) and the irradiance just below the surface (EPAR0-) from the measurement at 490 nm, in which 64% of acceptable quality profiles (for a more detailed explanation see supplementary material).

In the case of EPAR0-, it represents an instantaneous value. To get the daily light regime of phytoplankton, we integrated over the length of a day in the location of the float, following the procedure described in Mignot et al. (2018). At the end of this procedure, we got profiles of daily integrated irradiance, which were used for two purposes: (i) Compute the euphotic zone depth based on the depth of an isolume and (ii) to get the daily integrated irradiance received at the depth of the chlorophyll-*a* maxima along with the BGC-Argo float time series.

The euphotic zone was defined by the depth of the 0.04 molQ m$^{-2}$ day$^{-1}$ isolume. This light level is equivalent to an instantaneous PAR irradiance of 1 µmolQ m$^{-2}$ s$^{-1}$, assumed to be constant for 12 hours. The light level was chosen to be the mean compensation irradiance estimated for low-light adapted ecotypes of *Prochlorococcus* (Moore et al., 1995). Note that 12 hours assumption is near the daytime duration at the mean float position, and quite like the conditions used in laboratory experiments to assess *Prochlorococcus* photo-physiology.

The vertical distribution of chlorophyll-*a* in the Oxygen Minimum Zone of the ETNP frequently had two maxima, but profiles with a single maximum or with a more complex distribution are also found (Márquez-Artavia et al., 2019). All these cases were observed in the analysed dataset used in this work, and some representative profiles are shown in figure 2. Nonetheless, we focused on profiles with single or double maxima to analyse the evolution of the physical and biogeochemical conditions related to them.

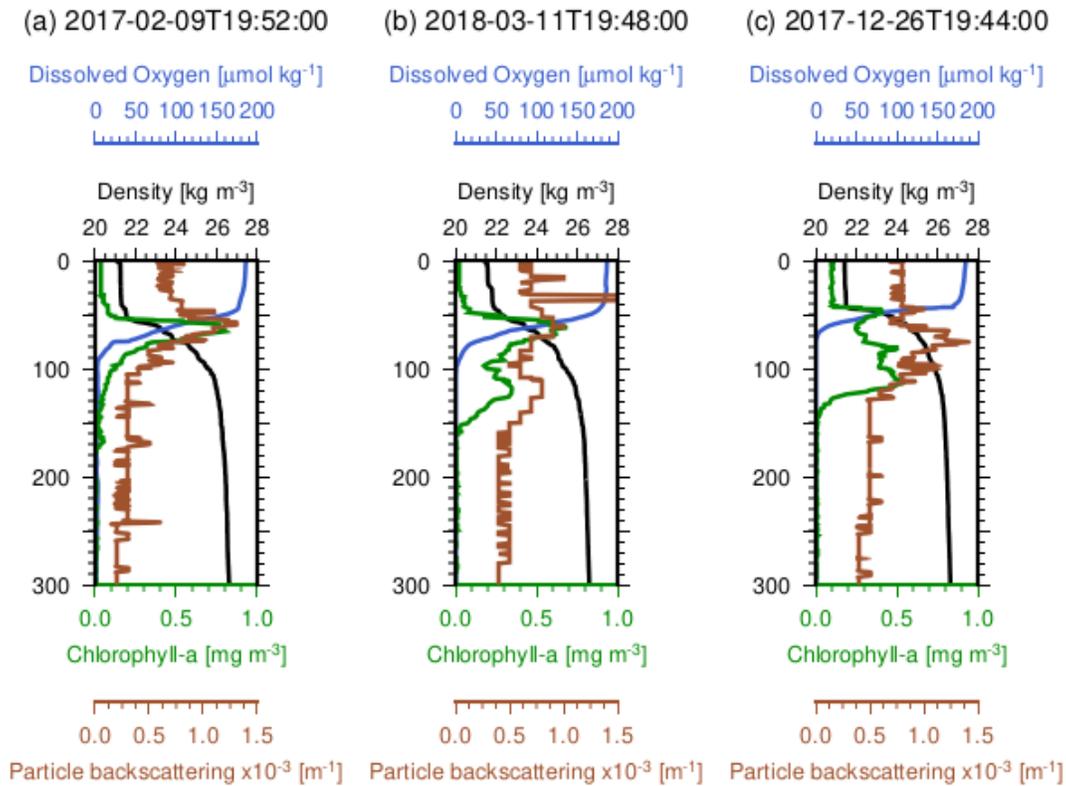

**Figure 2.** Represetative vertical profiles gathered by the BGC-Argo float. It is shown examples where the chlorophyll-a presented one (a), two (b) and three maxima (c). It is also shown the vertical distribution of of density, dissolved oxygen and particle backscattering (proxy of phytoplankton biomass). Note that chlorophyll-a and particle backscattering maxima coincided in depth.

We obtained the oceanographic conditions associated with the chlorophyll-*a* maxima considering the shape of the chlorophyll-*a* profile, which in most cases can be defined by a fitting function with one or two Gaussian curves (one for each maximum) plus an exponential decay. Similar approaches have been used elsewhere in the literature (Barbieux et al., 2018; Mignot et al., 2011; Muñoz-Anderson et al., 2015). We fitted the functions with the Levenberg-Marquardt algorithm, to get the parameters defining the depth and width of the chlorophyll-*a* maxima. The depth of each maximum corresponds to the center of the Gaussian curve, while the width covers the region between the center ±1 standard deviation. Properties like chlorophyll-*a*, and PAR, were represented by averaging the values within the width of each maximum. Linear correlations between variables were calculated using the Pearson's coefficient.

At each time step, the conditions related to each maximum are defined by the values of chlorophyll-*a* (chla), $b_{bp}700$ (as a proxy of carbon or phytoplankton biomass; $b_{bp}$), the depth of the nitracline (Z_nitracline), dissolved oxygen (DO), downwelling irradiance at PAR (PAR) and the ratio between $b_{bp}700$ and chlorophyll-*a*, which is a proxy of the carbon to the chlorophyll-*a* ratio ($b_{bp}$:chla). We used Z20 to identify how the conditions of the chlorophyll-*a* maxima were related to the thermocline/nitracline movements.

To find patterns of co-variability, and to decompose the temporal evolution of the chlorophyll-*a* maxima conditions in a set of uncorrelated modes, we made a Principal Component Analysis (PCA). The eigenvalues and eigenvectors were obtained by singular value decomposition of the standardized data. Individual observations were discarded if at least one of the variables did not have valid data. PCA was done using the package FactoMine version 1.42 in R 3.5.3 (Sébastien et al., 2008; R Core Team, 2019).

2.3 Long Rossby Wave Model

We used a simple analytical linear model to simulate the thermocline depth by the effects of LRWs, which considers both the local effects of wind stress on the thermocline depth, and the remote effects given by the propagation of perturbations (see Kessler 2006; Appendix). This simplified linear model has been used in several previous works, and it reproduces the propagation of Long baroclinic Rossby waves in tropical regions, explains the observed annual cycle of Z20 in the whole ETNP, and it has been useful to understand the local circulation of the Pacific off Mexico (Watanabe et al., 2016; Godínez et al., 2010; Kessler, 2006). Most of these previous works compared the model with observations, and given the correspondence between them, they agree that LRWs are a relevant physical process to explain the dynamic of the ETNP, and more specifically the low-frequency variability of the thermocline.

Following the approach of the previous works mentioned above, we used the model to compare the simulated thermocline depth (Z20) with the observations from BGC-Argo float at the ocean interior and with satellite altimeter measurements of SLA at the surface. These comparisons were made considering the baroclinic nature of LRWs, which means that the surface expression as SLA will be also observed as thermocline movements. In addition, we computed the Ekman pumping velocity to compare and cross-correlate with Z20. This will allow us to assess if the thermocline movements can be explained only by the local wind forcing. Ekman pumping velocity was computed daily in the dates sampled by the BGC-Argo float using the CCMP version 2 data product with a spatial resolution of 0.25°x0.25°, and it was interpolated linearly to the BGC-Argo float position. The highest correlation and lag are reported.

## 3. Results
### 3.1. Sea Level anomalies and the westward propagation of signals

The SLA spatial and temporal variability had three remarkable features around the BGC-Argo float locations. First, there was large scale process of annual period: During May 2017 the BGC-ARGO sampled within a region of positive values of SLA extending more than 8000 km in longitude (8-15 °N and 80-130 °W; Figure 3a), that changed to negative SLA after six months and with a similar spatial pattern (Figure 3b). Second, this signal appears to originate in the east and propagate westward, appearing as slanted features in the longitude-time diagrams of the SLA (Figure 3c). Third, the westward propagating signals showed an estimated phase speed of 13.1 ± 1.6 km day$^{-1}$ (Figure 3c), like the values reported near 13 °N by Chelton and Schlax (1996). In summary, there is an annual signal in the SLA with a westward propagation and phase speed that seems to have the characteristics of LRWs.

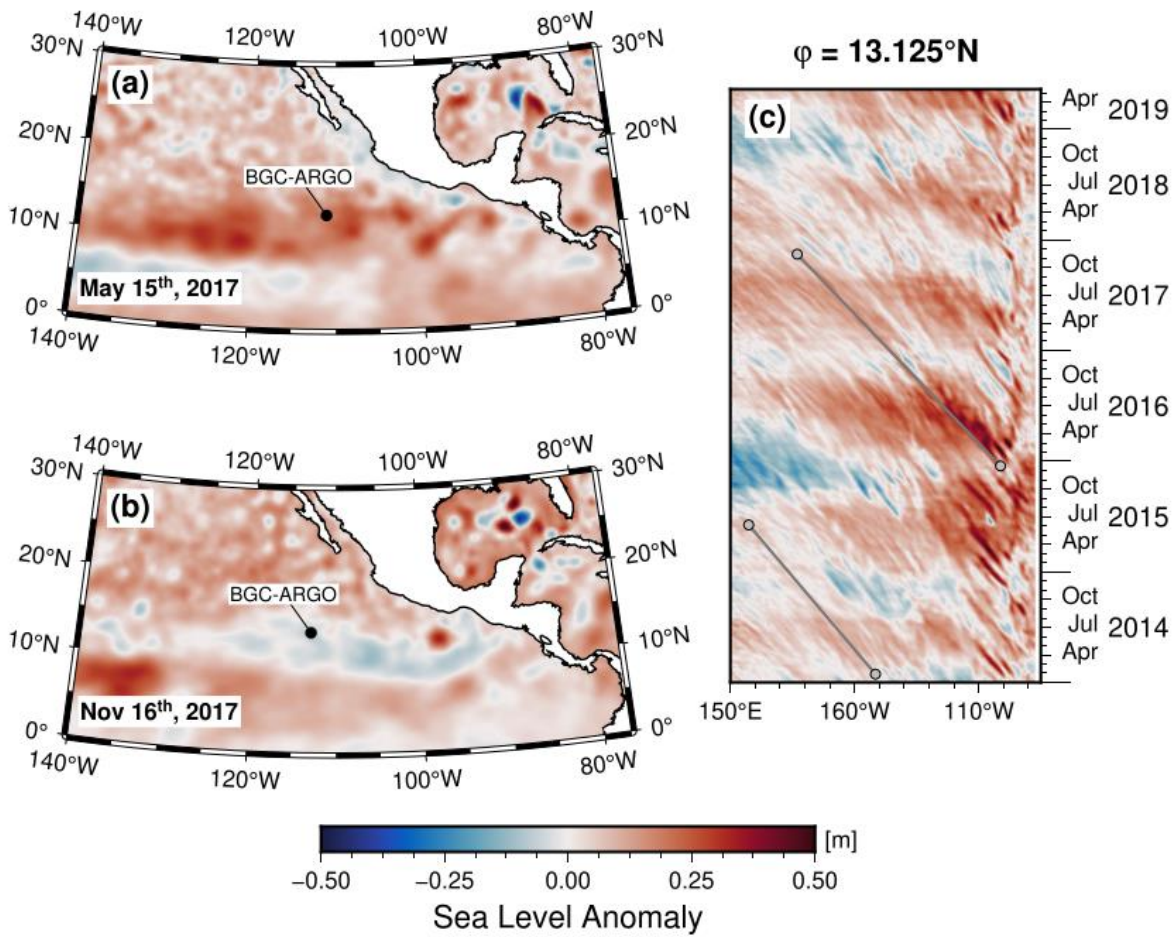

**Figure 3.** Spatial and temporal evolution of the Sea Level Anomaly (SLA) near the BGC-Argo float position. It is shown the spatial distribution of SLA on May 15th, 2017 (a) and November 16th, 2017 (b). A longitude-time plot at 13.125°N is shown in (c) with two examples of the lines used to estimate the phase speed of Rossby Waves.

Interestingly, the evolution of SLA and its LRW-like characteristics coincided with the temporal variability of chlorophyll-*a* presented in figure 4. Along with the time series it is possible to observe two persistent chlorophyll-*a* maxima change in depth over time. The chlorophyll-*a* maxima were deeper during the period of positive SLA (e.g. May 2017), while during the negative period (e.g. November 2017) they were shallower (Figure 4a and b). Both chlorophyll-*a* maxima simultaneously presented depth changes of ~40 m with a nearly annual signal and following the movements of Z20, the pycnocline, and nitracline (Figure 4b). It can be noted that the shallower maximum increased its chlorophyll-*a* content when the pycnocline, Z20, and nitracline move towards the surface, while the deeper maximum does not show strong changes in its pigment concentration during the record but it generally remains within the euphotic zone (Figure 4a).

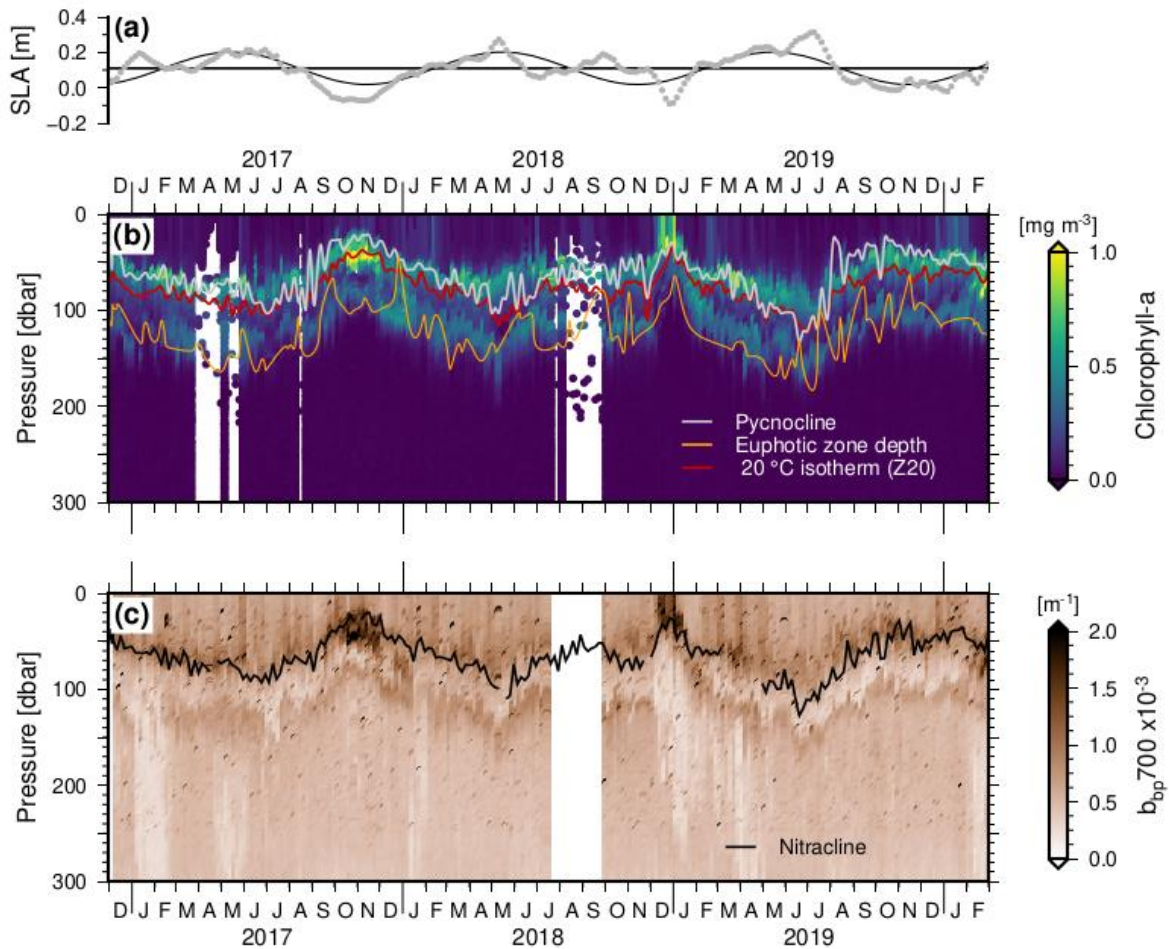

**Figure 4.** Temporal evolution of Sea Level Anomaly (SLA; a), chlorophyll-*a* (b) and the particle backscattering coefficient measured at 124° and 700 nm wavelength (bbp700). The position of the 20 °C isotherm (Z20), pycnocline, nitracline and the euphotic zone depth are also shown.

Given the chlorophyll-*a* maxima differs in their position in the water column, it is expected they presented differences in other properties. For example, we found the shallower maximum in more oxygenated waters (112.4 ± 31 µmolO$_2$ kg$^{-1}$) than the deeper chlorophyll-*a* maximum (0.8 ± 0.8 µmolO$_2$ kg$^{-1}$). Based on this difference in the dissolved oxygen concentration, we called each Chlorophyll-*a* Maximum (CM) as the oxic-CM and the suboxic-CM respectively. It is important to mention that we observed profiles with one or several maxima (Figure 2). However, they represent 3.1% of the good quality profiles analysed in this work, while the case with two maxima dominates, and it was found in 96.9% of the cases.

To discriminate if the chlorophyll-*a* maxima are produced by either phytoplankton growth enhancements or by a photoacclimation, we also showed the temporal evolution of the $b_{bp}700$ which is a proxy of organic carbon or phytoplankton biomass (Boss et al., 2015; Martinez-Vicente et al., 2013; Rasse et al., 2017; Stramski et al., 2004). Both the oxic and suboxic-CM coincided with maxima in the $b_{bp}700$ or biomass (Figure 2 and 4c). Chlorophyll-*a* and $b_{bp}700$ at each maximum presented a nearly annual signal in their depth displacements (Figure 4). The Person's correlation coefficients between the time series of chlorophyll-*a* and bbp700 were of 0.76 (n=153, df=151, p-value < 0.05) for the oxic-CM, and 0.89 (n=149, df=147, p-value < 0.05) for the suboxic-CM. The nearly annual oscillation was observed in other properties such as density and oxygen (Figure S1).

2 Chlorophyll-*a* on density coordinates and effects of LRWs

As shown above, the two persistent chlorophyll-*a* and $b_{bp}700$ maxima (oxic-CM and suboxic-CM) displayed depth variations simultaneously (Figure 4). However, in the density space, both maxima had a different response (Figure 5a). The oxic-CM (shallower and in less dense water) was sparsely distributed in the density space, and it moves between isopycnal surfaces through the year (Figure 5a). During May 2017, the oxic-CM was in the water with a density lower than 23 kg m$^{-3}$ at ~70 m depth, and with a thickness of ~20 m. Then, the oxic-CM moved to denser waters reaching the 25 kg m$^{-3}$ isopycnal during November 2017, when it became wider and shallower (Figures 4 and 5a). By contrast, the suboxic-CM had low variability in the density space, remaining near the 26 kg m$^{-3}$ isopycnal during the full record (Figure 5a).

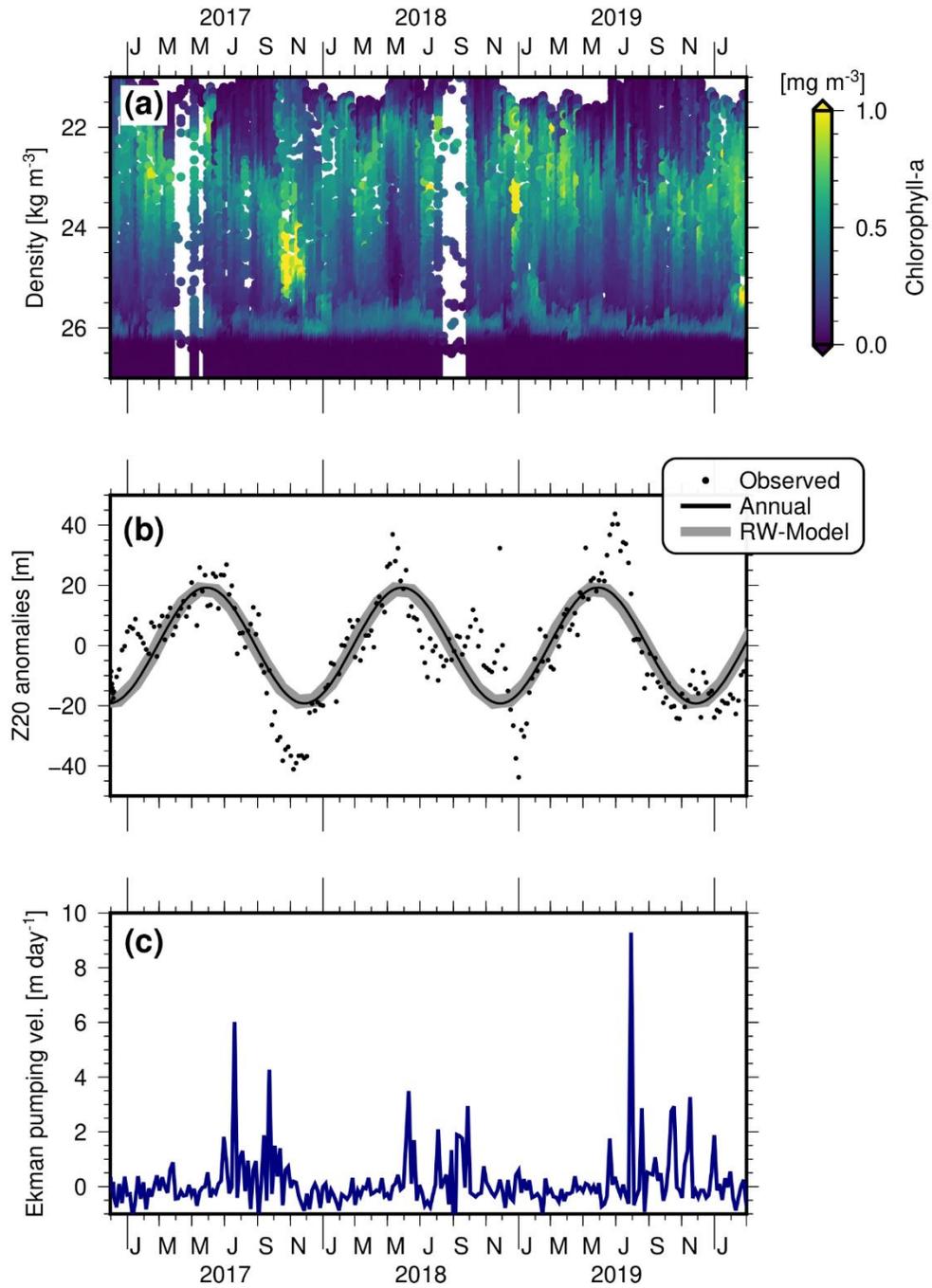

**Figure 5.** Temporal variability of chlorophyll-a (a), the depth of 20C isotherm; Z20 (b) and the Ekman pumping velocity. In (b) **is** depicted the BGC-float observations of Z20 (dots), their annual harmonic (black line) and the estimated Z20 from a theoretical Rossby wave model (gray thick line).

Based on the above results the response of each maximum to the pycnocline displacements is different, but they could be associated with the isopycnal movements caused by the propagation of LRWs. The LRWs are considered a potential mechanism considering that: (i) SLA spatial pattern suggests a large-scale process (Figure 3a and b), and also due to (ii) the nearly annual oscillation observed in the depth of the pycnocline (Figure 4). This annual oscillation of the pycnocline at 13 °N can be related to the propagation of LRWs as discussed in previous works (Kessler 1990 and 2006).

The analytical model presented in section 2 was then used to assess if the observed Z20 followed the theoretical prediction caused by the propagation of LRWs. It should be noted that the mean position of Z20 from the fitting procedure was $78.67 \pm 0.72$ m. The amplitude and phase of the annual harmonic were $19.25 \pm 1.01$ m and $2.32 \pm 0.05$ radians, respectively. This annual signal of Z20 is quite like the model results. Based on the observations, the annual harmonic accounts for 61.5 % of the total variance of Z20. The direct effect of the wind by the Ekman pumping can be considered as an alternative explanation to the temporal evolution of Z20. However, the comparison of the temporal variability of Ekman pumping and Z20 indicates that there is a lag between the timeseries. Note that the maximum Ekman pumping velocities occurred before than in Z20 (Figure 5b and c). Because of this we made a cross-correlation that indicates a maximum significant correlation when Z20 was lagged by 85 days ($r = -0.3$, p-value $< 0.05$, n=239). This phase difference indicates that a local response of Z20 to the wind cannot explain the observed temporal variability.

During the analysed record we observed that Z20 had strong departures from the annual signal and the predicted results gathered from the LRWs model (Figure 5b). For example, in October-November of 2017, the Z20 anomalies reached -40 m instead of the -20 m predicted by the annual harmonic and the model (Figure 5b). It means that thermocline was 20 m shallower than expected, and they are related to processes with frequencies higher than the annual. This implies that when the annual LRWs are in phase with other phenomena, it can cause strong Z20 movements in each direction causing these strong departures from the annual cycle.

Certainly, it is during these strong departures from the annual cycle that the chlorophyll-*a* increased substantially in the shallower chlorophyll-*a* maximum, being like bursts or anomalous events rather than a cyclic phenomenon (Figure 5). The annual cycle of the chlorophyll-*a* had an amplitude of 0.03 mg m$^{-3}$, accounting for the 4.6% of the total variance. The chlorophyll-*a* enhancements occurring in November 2017 and December 2018 (Figure 4a and 5a) had a concentration greater than 0.9 mg m$^{-3}$ and with a duration of 3 months, that in terms of standardized data, these short bursts of chlorophyll-*a* enhancements exceed the mean (0.34 ± 0.1 mg m$^{-3}$) by more than 3 standard deviations.

LRWs model can explain the low-frequency variability of Z20, but not the higher frequency variability. Here we first showed some information supporting the occurrence of LRWs of annual periods in the ETNP and analysis of eddy occurrence in the studied region. It can be noted the quite good correspondence of the spatial patterns of the annual cycle from sea level anomalies and the model which is presented in figure 6. Thus, it is supporting evidence to relate the low-frequency variability of the ETNP with the propagation of LRWs.

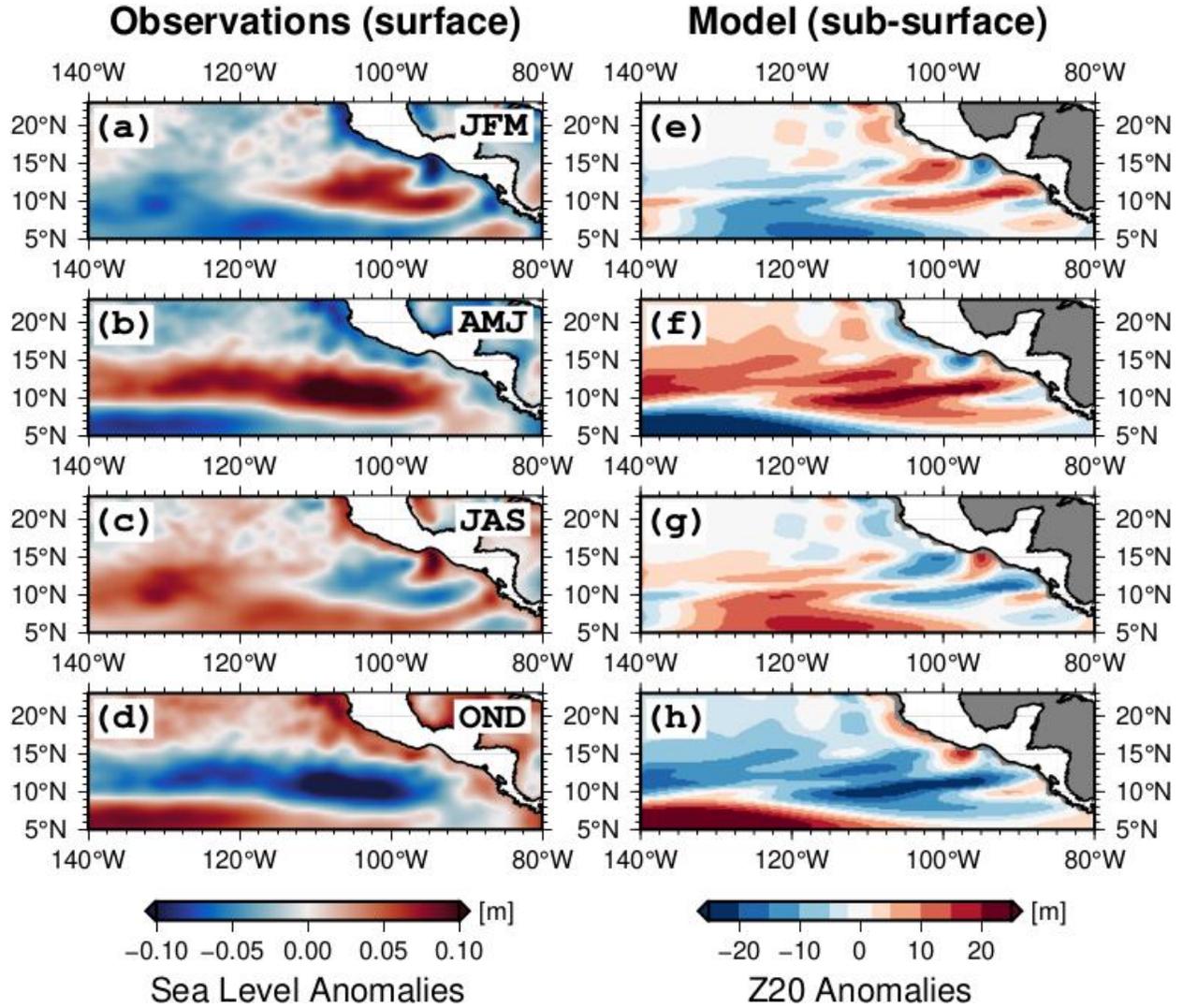

**Figure 6.** Comparison between the annual cycle from altimetry observations (a-d) and theoretically estimated from a Rossby wave model (e-h). The annual cycle of both datasets was quarterly averaged.

Secondly, the role of eddies on the departures from the annual cycle is explored in figure 7 where we classified the observations of SLA depending if the float was inside or outside eddies. For example, on January 15$^{th}$, 2017 the float was inside an anticyclone, the SLA was positive (Figure 7b), and hence the thermocline was deeper. The opposite case was recorded on December 31$^{st}$, 2018 during the influence of a cyclonic eddy (Figure 7c). Nonetheless, some events with remarkable differences from the annual cycle were not associated with eddies, as it can be noted on June 19$^{th}$, 2019 (Figure 7d). During this event, the float was within a region of relatively strong positive anomalies extending from 112-124 °W (~1200 km). As it can be seen in figure 7d, several eddies were detected inside this region of positive SLA, but the float was not within any of them or their edges (Figure 7a and d). Finally, this region with positive sea level anomalies, with a spatial extent of ~1200 km, originates at the coast and propagates westward (Supplementary animation).

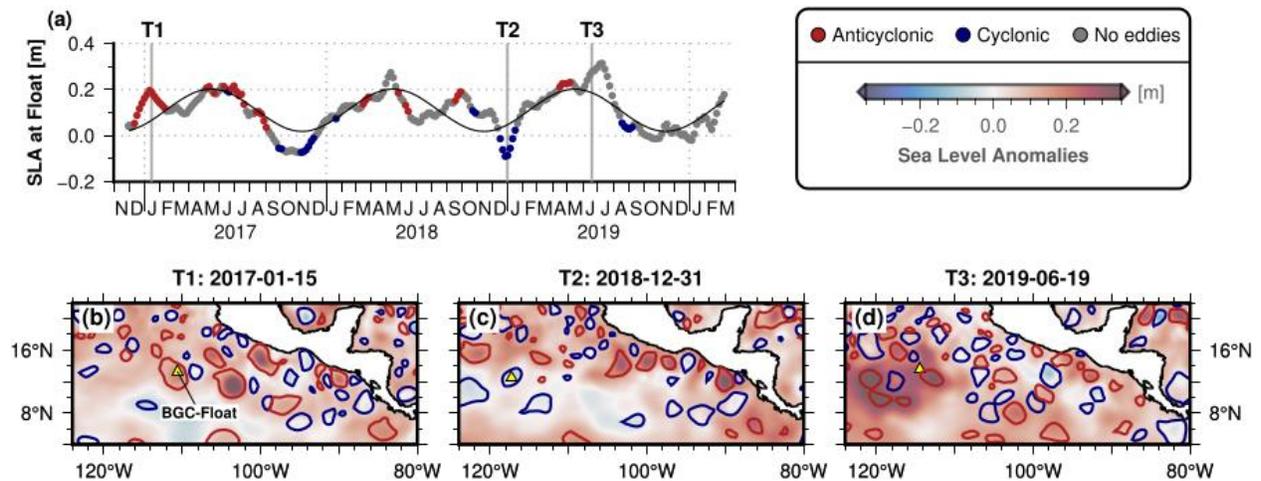

**Figure 7.** Temporal evolution of the Sea Level Anomaly (SLA) and occurrence of eddies on the float position. In (a) it is shown the SLA at the BGC-Argo float position. Each observation of SLA was classified and colored considering if the BGC-Argo float was outside of any Eddy (gray), inside anticyclonic (red points) or within cyclonic eddies (blue). The black line in (a) represents the annual cycle. Three temporal snapshots of the SLA (T1, T2, T3) are represented in the lower panels (b-d). The red and blue contours in the maps delineate spatial boundaries of anticyclonic and cyclonic eddies. We do not screened eddies by lifespan or size.

3 Principal Component Analysis (PCA)

We implemented a PCA to decompose the variability of the conditions at the depth of each chlorophyll-*a* maximum (the oxic-CM and suboxic-CM), and most importantly to look for patterns of co-variability between variables. The variables used were the dissolved oxygen (DO), nitracline depth (Z_nitracline), daily integrated PAR (PAR), chlorophyll-*a* (chla), $b_{bp}700$ ($b_{bp}$), and the ratio between $b_{bp}700$: and chlorophyll-*a* ($b_{bp}$:chla). The results of the PCA are presented as bi-plots (Gabriel, 1971; Legendre and Legendre, 2012) in figure 8 and the correlation coefficient of each variable and the statistical significance with a given principal component are shown in tables 1 and 2.

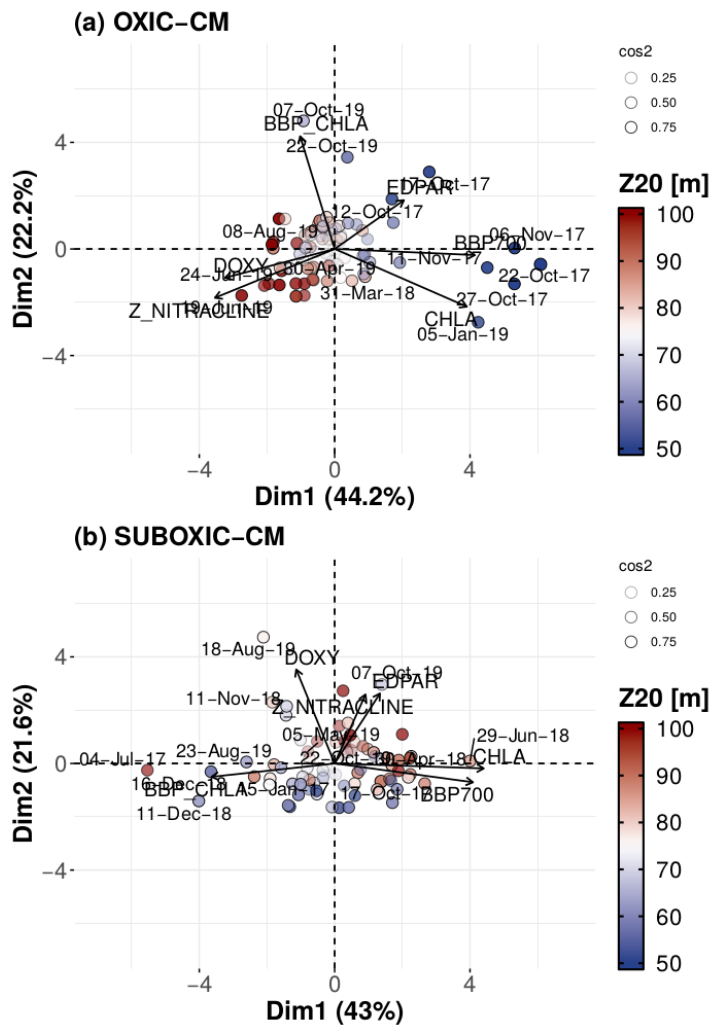

**Figure 8.** PCA bi-plots for the conditions associated with the oxic-CM (a) and the suboxic-CM (b). Each dot represents an individual observation projected into PCs, which are linear combinations of the original variables: chlorophyll-*a* (chl), particle backscattering coefficient (bbp; proxy of carbon), nitracline depth (z_nitracline), daily integrated irradiance at PAR band (PAR), dissolved oxygen (DO), and the ratio between bpp700 and chlorophyll-*a* (bbp:chl). Each dot is colored according to the vertical position of the 20 °C isotherm (Z20).

In the bi-plots, the individual observations are projected in a nondimensional space defined by the principal components (PCs). In our case, the coordinates for the individual observations in each PC, are obtained by the multiplication of the eigenvectors (loadings) by the square root of the corresponding eigenvalues. The contribution of each variable (Z_nitracline, chlorophyll-*a*, bbp700, etc) is represented using arrows. In general terms, the angle between the arrows and the PCs is related to its correlation. It can get values between 0° for a positive correlation and 180° for a negative association. An angle of 90° between an arrow and a given PC, indicates that there is no correlation. Variables with arrows in the same direction and with similar angles will have strong positive correlations. The arrow representation is done to try to find a physical/biological explanation for the mathematically constructed PCs.

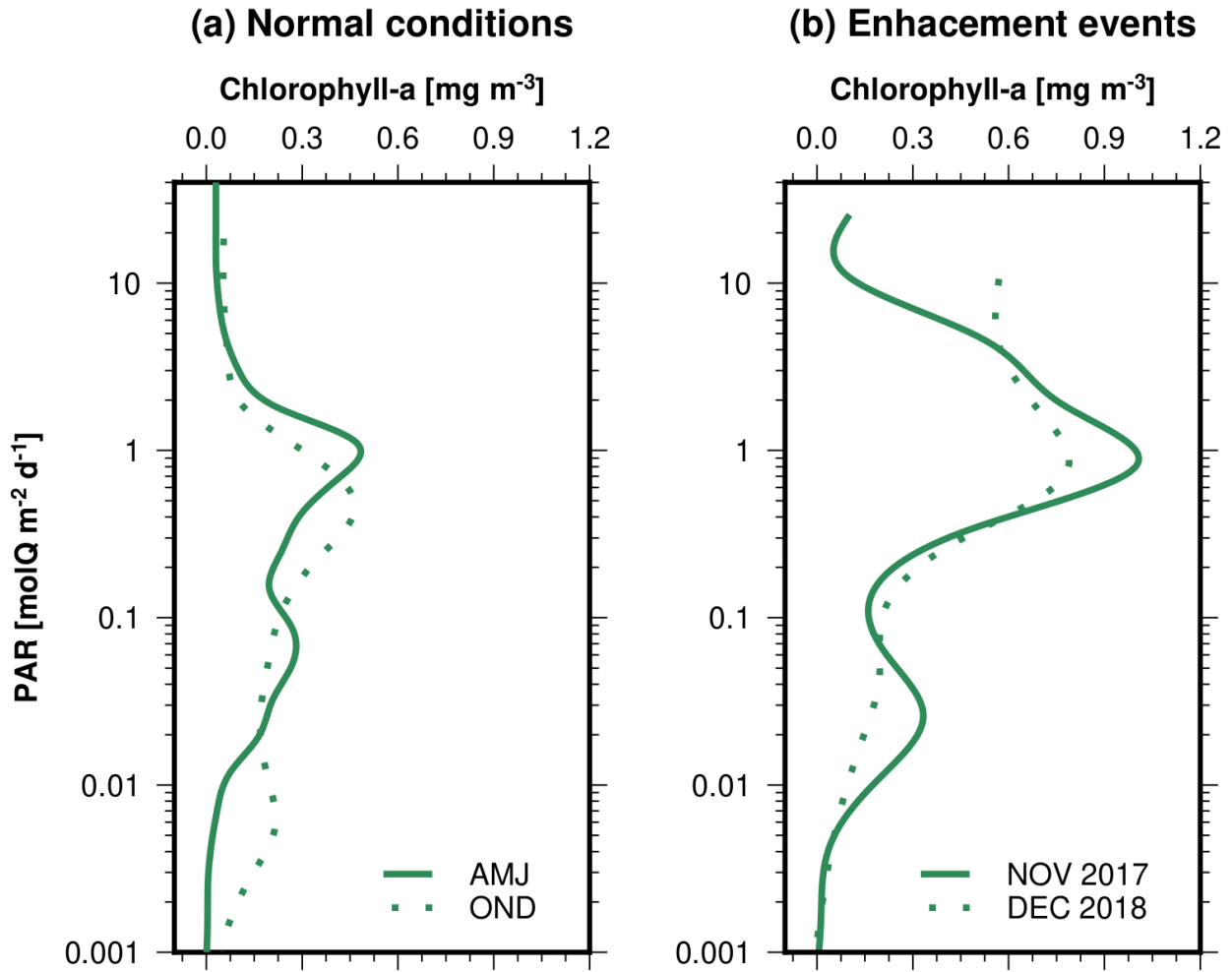

**Figure 9.** Distribution of chlorophyll-a as a function of daily integrated PAR. Chlorophyll-a profiles gathered on "normal" conditions between April-June (AMJ) and October-December (OND) are compared to represent differences along the seasonal cycle (a). The enhacements events observed during November 2017 and December of 2018 are represented in (b).

The first PC of the oxic-CM accounts for 44.2% of the variance and it is dominated by the contribution of $b_{bp}700$, chlorophyll-*a,* and the nitracline depth, which covaried (Figure 8a and Table 1). Along with this first PC, the observations with the highest (intense red points at the left) and lowest (intense blue points at the right) Z20 values were separated indicating that they are not seasonal events (Figure 8a). The second PC for the oxic-CM (22.2 % of the total variance), had the major contributions given by the ratio $b_{bp}700$:chla, which were positively correlated with this axis (Figure 8a). Chlorophyll-*a* had a negative correlation with the second PC (Table 1). It indicates that the second mode could be attributed to photoacclimation (chlorophyll-*a* increases when PAR decreases).

The pattern found after doing the PCA with the conditions of the suboxic-CM was different (Figure 8b). In the first PC Chlorophyll-*a*, $b_{bp}700$, and the ratio $b_{bp}700$:chla co-varied (36.6 % of the total variance) but, they were nearly orthogonal to the changes in PAR and nitracline depth, which were associated significantly with the second PC (Figure 8b and Table 2). Oxygen had a significant correlation with PC1 although it was low (Table 2).

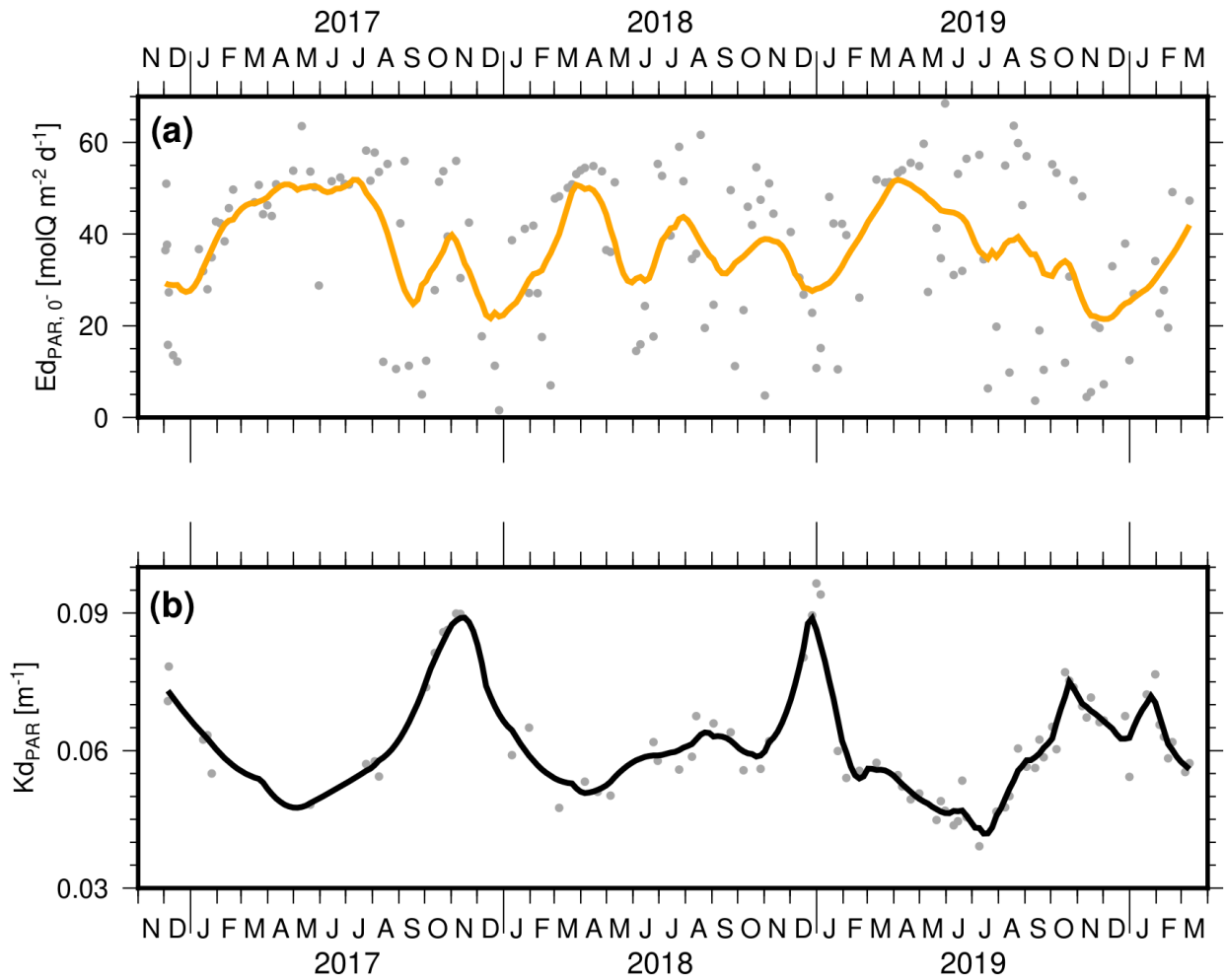

**Figure 10.** Time series of daily integrated PAR (a) and the diffuse attenuation coefficient (b) from the BGC-Argo float measurements. Grey points are observations that were smoothed by using a LOESS regression and represented with a solid line in both panels.

The PCA done with the conditions of the oxic-CM suggest that the depth of the nitracline or the nutrient concentration is the main factor causing the chlorophyll-*a* and biomass enhancements, while the light seems to have a limited effect. This behaviour differs from those observed in oligotrophic regions, where irradiance drives seasonal changes in the chlorophyll-*a* maximum, which follows the movements of isolumes. For this reason, we plotted the chlorophyll-*a* as a function of PAR (Figure 9). We first compared the mean chlorophyll-*a* profiles gathered during April-June and from October-December. This represents two contrasting conditions in which the chlorophyll-*a* maxima reach their deepest and shallowest positions along the seasonal cycle. The chlorophyll-*a* during these contrasting conditions was quite similar, although the PAR changed remarkably with the maximum during April-June and the minimum in October-December (Figure 9a). In the case of the enhancement events, the PAR does not exceed the 1 molQ $m^{-2}$ $d^{-1}$ at the center of the oxic-CM but it seems to cover a wider range of PAR values than in normal conditions (Figure 9b).

## 4. Discussion

The analysis presented here is supported by three years of measurements obtained by a BGC-ARGO. This kind of instrumentation is providing an unprecedented view of relevant biogeochemical processes, including those regulating chlorophyll-*a* and phytoplankton growth in the world oceans (e.g. Barbieux et al., 2018; Mignot et al., 2014). More specifically by using BGC-ARGO floats in the Arabian Sea OMZ, it has been found two chlorophyll-*a* maxima which can be regulated by propagating LRWs (Ravichandran et al., 2012). However, in the ETNP this is the first time that there is a sufficiently long time series to resolve those processes and, at the same time, a data set that allows studying the role of light and nutrients in the dynamic of the phytoplankton communities. Here we will discuss the effects of physical phenomena on the vertical distribution of properties, and then we will deal with the implications for the chlorophyll-*a* variability.

We can consider several physical phenomena to explain the thermocline movements in an annual cycle. For example, mesoscale eddies, the local Ekman pumping alone, and the propagation of LRWs. Mesoscale eddies can be easily discarded to explain the low-frequency variability because it is not expected to cause an annual signal, just as it has been demonstrated in figure 7. On the other hand, if the thermocline responds to the wind local forcing directly, we will expect a high correlation with zero or short lag. However, that was not the case and the thermocline was lagged by 2.8 months from the Ekman pumping velocity, supporting the idea that thermocline responds to a remote forcing as previously suggested by Fiedler and Talley (2006). The remote forcing can be the propagation of LRWs whose occurrence in the ETNP is supported by the large scale pattern in SLA (Figure 3a and b), the westward propagating signals at phase speeds characterising LRWs (Figure 3c), and also given the correspondence between the theoretical LRWs model and observations of Z20 and SLA (Figure 5 and 6). The occurrence of annual Rossby waves and their effect on the thermocline low-frequency variability is not new, and it has been documented extensively in previous works (Capotondi et al., 2003; Chelton and Schlax, 1996; Kessler, 1990; Meyers, 1979; Polito and Liu, 2003; Qiu et al., 1997). Hence, at 13 °N the thermocline/nitracline annual cycle is driven by the propagation of LRW forced by the wind stress curl as a cumulative forcing from the eastern coast to the open ocean, rather than a direct effect of the wind by Ekman pumping alone.

The annual cycle of the thermocline/nitracline accounts for a considerable portion of the total variance, but other events at lower periods (~4 months) were observed and they caused remarkable differences from the annual cycle (Figure 5b). Again, we can attribute these events to several processes. We first assessed the mesoscale eddies because the BGC-Argo float was near the path of eddies generating in Tehuantepec by the Central America wind jets (Aleynik et al., 2017; Chelton et al., 2007). Our analysis showed that several eddies pass over the BGC-Argo float (Figure 7), and some of them seem to cause strong departures from the annual cycle. Nonetheless, other events were not associated with eddies such as the event occurring on June 19$^{th}$, 2019. In this case, strong departures from the annual cycle correspond to a larger scale process than eddies (Figure 7d), which originates at the coast and propagates westward (Supplementary material). Given these characteristics, the phenomenon could be associated with Rossby waves of shorter periods than annual that had been reported to occur in the ETNP (Polito and Liu, 2003). Actually, it seems that eddies propagate along with these Rossby waves as has been demonstrated previously (see supplementary animation; Polito and Sato, 2015). Summarizing, the propagation of LRWs of annual period generates a "*background*" signal over the thermocline/nitracline that is modulated by higher frequency processes including, but not limited to, mesoscale eddies.

We have shown that the thermocline/nitracline and isopycnals are affected by the occurrence of several processes. Here, we are going to focus on the implications that it has on the chlorophyll-*a* field. Firstly, in the analysed dataset from the BGC-float, the vertical distribution of chlorophyll-*a* presented two persistent maxima, which coincided with a local increase in the bbp700 signal; a proxy of the particulate organic carbon or phytoplankton biomass (Boss et al., 2015; Stramski et al., 2008). Consequently, the chlorophyll-*a* maxima can be interpreted as phytoplankton biomass maxima with a remarkable depth variability (Figure 4). The vertical displacements of the maxima were synchronized, following the evolution of SLA, the temporal variability of the thermocline/nitracline (Figure 4).

Because both Rossby waves and eddies modify isopycnals' position in the water column, they can also change the vertical distributions of nutrients, and consequently the chlorophyll-*a*.

However, the isopycnals movements can affect the chlorophyll-*a* by either a mechanical action if the phytoplankton is linked to a given isopycnal surface, or as a biological response of phytoplankton to changes in nutrients and light. Thus, if chlorophyll-*a* is plotted in the density space (Figure 5a), where the effects of circulation are removed, the chlorophyll-*a* maxima will look flat over time if only a mechanical action occurs in response to the isopycnals movements. It is noticeable that the shallower chlorophyll-*a* maximum, which we called the oxic-CM, oscillated in the density space. In contrast the deeper maximum or the suboxic-CM, remains tightly coupled to the 26 kg m$^{-3}$ isopycnal. The oxic-CM intensifies when it moves to denser waters, possibly as a response to an increase in the nutrient supply (Figure 5a). By contrast, the suboxic-CM is only mechanically displaced (Figure 5a). Hence, each maximum has a different response to the isopycnals movements.

A more detailed view of the factors controlling the evolution of each chlorophyll-*a* maximum is based on the interpretation of the PCA (Figure 8). In the case of the oxic-CM, the first PC was dominated by the co-variability between chlorophyll-*a*, bbp700, and the nitracline depth variability, supporting the idea that the changes in chlorophyll-*a* are controlled by the nitrate supply (Figures 4 and 8). Given the above observations and considering that the oxic-CM is located near the nitracline, we considered that it fits in the typical stable water structure proposed by Cullen (2015) from meso and eutrophic conditions, in which the movements of the nitracline control the chlorophyll-*a* concentration, the biomass, and the primary production.

The above explanation implies that the chlorophyll-*a* variability of the oxic-CM is driven by nutrients and not by light. Light-driven cycles are common to occur in oligotrophic environments. For example in the North Pacific subtropical gyre the chlorophyll-*a* maximum follows the depth of the isolumes, but it has a seasonal cycle in the chlorophyll-*a* concentration: it becomes more intense during the summer when light penetrates deeper in the water column, while during the winter the chlorophyll-*a* decreases because light limitation precludes the use of nutrients even if they are available (Letelier et al., 2004). By contrast in our studied region the seasonal cycle of the chlorophyll-*a* at the oxic-CM had a very low amplitude accounting for a low percentage of the total variance (< 5%), and the chlorophyll-*a* maxima do not follow a given isolume (Figure 9).

The chlorophyll-*a* enhancements occurred at higher PAR values than the expected mean, but without exceeding the maximum mean value observed during April-June (~1 molQ $m^{-2}$ $day^{-1}$). During the enhancements events the chlorophyll-*a* increased by a factor of 4, and this is difficult to explain only considering the observed PAR variability. Hence the oxic-CM dynamics differentiate from those of oligotrophic conditions that are seasonally light driven. Instead of that, it fits into the typically stable water structure scenario, in which chlorophyll-*a* is controlled primarily by the nutrient supply (Cullen 2015).

Here we will discuss the second PC of the oxic-CM, which was dominated by changes in the $b_{bp}$:chla ratio, and presented a weak negative correlation between chlorophyll-*a* and PAR. This could be attributed to photoacclimation, but the profiles of chlorophyll-*a* as a function of PAR indicate that the chlorophyll-*a* concentration does not change despite the variability in light in seasonally contrasting conditions (Figure 9a). Hence, we do not find strong evidence of photoacclimation, and the second PC could be associated with changes in the phytoplankton community composition. This is difficult to assess with the current datasets, but it can be done in future works about the phytoplankton of the ETNP.

According to our analysis (Figure 8 and Table 1), the dissolved oxygen is another variable that seems to be related to the chlorophyll-*a* enhancements of the oxic-CM. In general terms the oxic-CM intensifies considerably when denser and deeper waters move toward the surface. Given the strong oxygen gradients of the OMZ, it implies a considerable decrease in the dissolved oxygen, changing from a maximum of 187.3 µmol $O_2$ $kg^{-1}$ during "*normal conditions*" to a minimum of 20.6 µmol $O_2$ $kg^{-1}$ when enhancements occur. The strong decrease in oxygen could limit the distribution and abundance of organisms at the depth of the oxic-CM, and consequently reduce the grazing pressure; a source of phytoplankton cell loss. These changes in phytoplankton loss have been considered as relevant factors affecting phytoplankton accumulation (Behrenfeld and Boss 2014, Arteaga et al., 2020). Here it seems to be not the primary cause, but it could contribute to some degree.

The suboxic-CM can be considered as a special case and it shows a different pattern that does not fit the typical stable water structure. It was located below the nitracline at 118 m depth, where the oxygen concentrations were below 5 µmol $kg^{-1}$, and the mean PAR was 0.26 molQ $m^{-2}$ $d^{-1}$ (0.6% of irradiance just below the sea surface); a set of conditions favouring the growth of low light-adapted ecotypes of *Prochlorococcus* in the OMZs (Garcia-Robledo et al., 2017; Goericke et al., 2000; Lavin et al., 2010).

In the multivariate analysis of the suboxic-CM conditions, the chlorophyll-*a*, bbp700, and the ratio bbp700: chla were the dominant variables in the first component, and they were nearly orthogonal (~ 90° angle) with the second principal component, where PAR and nitracline depth dominate the variability (Figure 8b). Thus, the chlorophyll-*a* and bbp700 do not co-varied with nitrate and light. Oxygen has a low correlation with the second component, and there was not a clear pattern that explains the variability of chlorophyll-*a* or bbp700 in the suboxic-CM. We considered that other variables such as biological interactions between *Prochlorococcus* with other bacteria of the OMZ could be responsible for the maintenance of the suboxic-CM, but it should be further assessed.

Finally, we highlight the persistence of the suboxic-CM observed in the BGC-float data set, which was present in 96.9% of all profiles (Figure 2). This persistence was also reported in the Pacific off Mexico between 16-21°N (Márquez-Artavia et al., 2019), and here it is corroborated for a more extended region of the ETNP. In contrast, the suboxic-CM seems to occur sporadically in the Arabian Sea, and in the eastern tropical South Pacific (Whitmire et al., 2009; Ravichandran et al., 2012; Wojtasiewicz et al., 2018). The reasons by which the suboxic-CM persists in the ETNP are not well understood, but it demonstrates that there are differences in the dynamics of the OMZs of the global ocean.

Future studies on the process affecting the phytoplankton communities of the OMZ are required. There is, therefore, a definite need for more instruments to measure regularly in the OMZs. For example, there is a need for directly measuring the nitrate concentration to shed more light on the use of this nutrient by the phytoplankton populations, especially in the suboxic waters of the OMZ.

## 5. Conclusions

The findings of this work agree with previous publications regarding the relevant contribution of annual LRWs to the thermocline/nitracline depth variability, and consequently on the isopycnals movements. However, we also found the occurrence of other physical processes of higher frequencies than annual (e.g. mesoscale eddies and Rossby waves of shorter periods), which modulate the background annual signal. It seems that a constructive interaction between all processes can produce extraordinary events, when the nitracline becomes very shallow, and increasing the nutrient supply into the euphotic zone. These kinds of extraordinary events are expected to occur more frequently during October-December when annual LRW displace the nitracline toward the surface along with the annual cycle.

The response of each chlorophyll-*a* maximum (the oxic-CM and suboxic-CM) to the vertical displacements of the isopycnals was different. The oxic-CM seems to be affected primarily by an increase in nutrient concentration and possibly by a diminished grazing pressure due to the strong changes in dissolved oxygen. This was observed during short bursts of intensification of both chlorophyll-*a* and biomass ($b_{bp}$) suggesting that are events of increased phytoplankton growth. A second-order variation in the oxic-CM, which is dominated by changes in the $b_{bp}$:chla ratio can be related to changes in the phytoplankton community composition because we do not find evidence of photoacclimation as a response to the seasonal changes in irradiance. Finally, the deeper chlorophyll-*a* maximum which we called the suboxic-CM, was persistent at a depth where PAR can sustain a net growth of *Prochlorococcus* low-light adapted ecotypes. This chlorophyll-*a* was only mechanically displaced along with the isopycnals and its temporal evolution seems to be decoupled from irradiance and nitrate variability.

## 6. Appendix: The Linear Long Rossby Wave Model

The linear model simulating the thermocline depth variability by the effect of LRW, has a solution with two terms: (i) the wind component ($h_W$) associated to the wind stress curl forcing ($\tau$) and, (ii) LRWs radiating from the eastern boundary ($h_B$) due to the variability of the thermocline depth near to the coast ($h_E$). Solution ii is important only near the coast and quickly decayed in the open ocean.

$$h_W(x,t) = \frac{-1}{c_r} \int_{x_E}^{x} e^{\frac{-R}{c_r}(x-x')} \nabla \times \left[ \tau \frac{\left(x', t - \frac{x-x'}{c_r}\right)}{f\rho} \right] dx' \text{ (i)}$$

$$h_B(x,t) = e^{\frac{-R}{c_r}(x-x_E)} h_E\left(t - \frac{x-x_E}{c_r}\right) \text{ (ii)}$$

Where $c_r$ denotes phase speed of LRWs ($-\beta c^2 / f^2$), R is the damping timescale, $f$ is the Coriolis parameter and $\rho$ is considered as a mean ocean density (1027 kg m$^{-3}$). The value for the gravitational waves speed was $c = 2 \frac{m}{s}$. The lower limit of the integration, $x_E$, is the longitude at the eastern boundary and the values of all parameters are the same used previously by Kessler (2016) for consistency. The boundary condition ($h_B$) was the annual cycle of the thermocline, averaged in the first four degrees from the coast.

We used wind data from the Cross-Calibrated Multiplatform version 2 (CCMPv2), between the period 2010-2018. From this data, the annual cycle of the wind stress curl has been estimated. The wind dataset has a spatial resolution of 0.25°x0.25° and it was acquired from Remote Sensing Systems (www.remss.com). We used the EN4 gridded dataset (Good et al., 2013) to compute the boundary condition ($h_B$). EN4 dataset is available at the Met office-UK website (https://www.metoffice.gov.uk/hadobs/en4/).

## 7. Acknowledgments

AM acknowledges the financial support received from CONACyT to develop this work as research assistant no. 19337. EB was supported by the project Dinámica y Termodinámica de la Corriente Occidental Mexicana (CICESE 612-691) and LSV acknowledge to Instituto Politécnico Nacional (SIP-IPN 20200669), and CONACyT for their support during the sabbatical.

The data from the floats were collected and made freely available by the international Argo project and the national programs that contribute to it (http://doi.org/10.17882/42182). We are indebted to the Principal Investigators of the Argo program and also acknowledged the availability of SLA, wind data, and the EN4.2.1 datasets available from Copernicus Marine Service (http://marine.copernicus.eu), Remote Sensing Systems (www.remss.com) and Met Office-UK

(https://www.metoffice.gov.uk/hadobs/en4/). The authors are grateful to Dr. William Kessler for his suggestion to represent chlorophyll on density space. We are also extremely grateful for the insightful comments given by Professor Paulo Polito and the anonymous reviewers that substantially improved this work.## 8. References

Aleynik, D., Inall, M.E., Dale, A., Vink, A., 2017. Impact of remotely generated eddies on plume dispersion at abyssal mining sites in the Pacific. Sci. Rep. 7, 1–14. https://doi.org/10.1038/s41598-017-16912-2

Antoine, D., André, J.-M., Morel, A., 1996. Oceanic primary production 2. Estimation at global scale from satellite (coastal zone color scanner) chlorophyll. Global Biogeochem. Cycles 10, 57–69.

Arteaga, L. A., Boss, E., Behrenfeld, M. J., Westberry, T. K., Sarmiento J. L., 2020. Seasonal modulation of phytoplankton biomass in the Southern Ocean. Nat Commun. 11:5364. https://doi.org/10.1038/s41467-020-19157

Barbieux, M., Uitz, J., Bricaud, A., Organelli, E., Poteau, A., Schmechtig, C., Gentili, B., Obolensky, G., Leymarie, E., Penkerc'h, C., D'Ortenzio, F., Claustre, H., 2018. Assessing the Variability in the Relationship Between the Particulate Backscattering Coefficient and the Chlorophyll a Concentration From a Global Biogeochemical-Argo Database. J. Geophys. Res. Ocean. 123, 1229–1250. https://doi.org/10.1002/2017JC013030

Barron, C.N., Kara, A.B., Jacobs, G.A., 2009. Objective estimates of westward Rossby wave and eddy propagation from sea surface height analyses. J. Geophys. Res. Ocean. 114, 1–18. https://doi.org/10.1029/2008JC005044

Behrenfeld, M. J., Boss, E. S., 2014, Resurrecting the Ecological Underpinnings of Ocean39

Plankton Blooms. Ann. Rev. Mar. Sci. 6, 167-194. https://doi.org/10.1146/annurev-marine-052913-021325

Belonenko, T. V., Bashmachnikov, I.L., Kubryakov, A.A., 2018. Horizontal advection of temperature and salinity by Rossby waves in the North Pacific. Int. J. Remote Sens. 39, 2177–2188. https://doi.org/10.1080/01431161.2017.1420932

Bittig, H., Wong, A., Plant, J., CORIOLIS-ADMT, 2018. BGC-Argo synthetic profile file processing and format on Coriolis GDAC. https://doi.org/http://dx.doi.org/10.13155/55637

Boss, E., Pegau, W.S., 2001. Relationship of light scattering at an angle in the backward direction to the backscattering coefficient. Appl. Opt. 40, 5503. https://doi.org/10.1364/ao.40.005503

Boss, E., Stramski, D., Bergmann, T., Scott Pegau, W., Lewis, M., 2015. Why should we measure the optical backscattering coefficient? Oceanography 17, 44–49. https://doi.org/10.5670/oceanog.2004.46.

Boss, E.S., Häentjens, N., 2016. Primer regarding measurements of chlorophyll fluorescence and the backscattering coefficient with WETLabs FLBB on profiling floats. New Jersey.

Briggs, N., Perry, M.J., Cetinić, I., Lee, C., D'Asaro, E., Gray, A.M., Rehm, E., 2011. High-resolution observations of aggregate flux during a sub-polar North Atlantic spring bloom. Deep. Res. Part I Oceanogr. Res. Pap. 58, 1031–1039. https://doi.org/10.1016/j.dsr.2011.07.007

Capotondi, A., Alexander, M.A., Deser, C., 2003. Why are there Rossby wave maxima in the Pacific at 10 degrees S and 13 degrees N? J. Phys. Oceanogr. 33, 1549–1563.

Chelton, D.B., Schlax, M.G., 1996. Global observations of oceanic Rossby waves. Science (80-. ). 272, 234–238. https://doi.org/10.1126/science.272.5259.234

Chelton, D.B., Schlax, M.G., Samelson, R.M., de Szoeke, R.A., 2007. Global observations of large oceanic eddies. Geophys. Res. Lett. 34, 1–5. https://doi.org/10.1029/2007GL030812

Cullen, J.J., 2015. Subsurface chlorophyll maximum layers: enduring enigma or mystery solved? Ann. Rev. Mar. Sci. 7, 207–39. https://doi.org/10.1146/annurev-marine-010213-135111

Fiedler, P.C., Talley, L.D., 2006. Hydrography of the eastern tropical Pacific: A review. Prog. Oceanogr. 69, 143–180. https://doi.org/10.1016/j.pocean.2006.03.008

Gabriel, K.R., 1971. The biplot graphic display of matrices with application to principal component analysis. Biometrika 58, 453–467. https://doi.org/doi:10.1093/biomet/58.3.453

Garcia-Robledo, E., Padilla, C.C., Aldunate, M., Stewart, F.J., Ulloa, O., Paulmier, A., Gregori, G., Revsbech, N.P., 2017. Cryptic oxygen cycling in anoxic marine zones. Proc. Natl. Acad. Sci. 114, 8319–8324. https://doi.org/10.1073/pnas.1619844114

Glatt, I., Dörnbrack, A., Jones, S., Keller, J., Martius, O., Müller, A., Peters, D. H. W. and Wirth, V., 2011. Utility of Hovmöller diagrams to diagnose Rossby wave trains. Tellus A. 63, 991-1006. https://doi.org/10.1111/j.1600-0870.2011.00541.x

Godínez, V.M., Beier, E., Lavín, M.F., Kurczyn, J.A., 2010. Circulation at the entrance of the Gulf of California from satellite altimeter and hydrographic observations. J. Geophys. Res. Ocean. 115, 1–15. https://doi.org/10.1029/2009JC005705

Goericke, R., Olson, R.J., Shalapyonok, A., 2000. A novel niche for Prochlorococcus sp. in the low-light suboxic environments in the Arabian Sea and the Eastern Tropical North Pacific. Deep. Res. I 47, 1183–1205. https://doi.org/10.1016/S0967-0637(99)00108-9

Good, S.A., Martin, M.J., Rayner, N.A., 2013. EN4: Quality controlled ocean temperature and salinity profiles and monthly objective analyses with uncertainty estimates. J. Geophys. Res. Ocean. 118, 6704–6716. https://doi.org/10.1002/2013JC009067
41

Kessler, W.S., 2006. The circulation of the Eastern Tropical Pacific: A review. Prog. Oceanogr. 69, 181–217. https://doi.org/10.1016/j.pocean.2006.03.009

Kessler, W.S., 1990. Observations of Long Rossby Waves in the Northern Tropical Pacific. J. Geophys. Res. 95, 5183–5217.

Killworth, P.D., Cipollini, P., Uz, B.M., Blundell, J.R., 2004. Physical and biological mechanisms for planetary waves observed in satellite-derived chlorophyll. J. Geophys. Res. C Ocean. 109, 1–18. https://doi.org/10.1029/2003JC001768

Kurian, J., Colas, F., Capet, X., McWilliams, J.C., Chelton, D.B., 2011. Eddy properties in the California Current System. J. Geophys. Res. 116, C08027. https://doi.org/10.1029/2010jc006895

Lavin, P., González, B., Santibáñez, J.F., Scanlan, D.J., Ulloa, O., 2010. Novel lineages of Prochlorococcus thrive within the oxygen minimum zone of the eastern tropical South Pacific. Environ. Microbiol. Rep. 2, 728–738. https://doi.org/10.1111/j.1758-2229.2010.00167.x

Legendre, P., Legendre, L., 2012. Numerical Ecology, Third edit. ed. Elsevier, Oxford, UK.

Letelier, R.M., Karl, D.M., Abbott, M.R., Bidigare, R.R., 2004. Light driven seasonal patterns of chlorophyll and nitrate in the lower euphotic zone of the North Pacific Subtropical Gyre. Limnol. Oceanogr. 49, 508–519. https://doi.org/10.4319/lo.2004.49.2.0508

Márquez-Artavia, A., Sánchez-Velasco, L., Barton, E.D., Paulmier, A., Santamaría-Del-Ángel, E., Beier, E., 2019. A suboxic chlorophyll-a maximum persists within the Pacific oxygen minimum zone off Mexico. Deep. Res. Part II Top. Stud. Oceanogr. 169–170. https://doi.org/10.1016/j.dsr2.2019.104686

Martinez-Vicente, V., Dall'Olmo, G., Tarran, G., Boss, E., Sathyendranath, S., 2013. Optical

backscattering is correlated with phytoplankton carbon across the Atlantic Ocean. Geophys. Res. Lett. 40, 1154–1158. https://doi.org/10.1002/grl.50252

Mason, E., Pascual, A., McWilliams, J.C., 2014. A new sea surface height-based code for oceanic mesoscale eddy tracking. J. Atmos. Ocean. Technol. 31, 1181–1188. https://doi.org/10.1175/JTECH-D-14-00019.1

Mcdougall, T.J., Barker, P.M., 2017. Getting started with TEOS-10 and the Gibbs Seawater (GSW) Oceanographic Toolbox.

McGillicuddy, D.J., 2016. Mechanisms of Physical-Biological-Biogeochemical Interaction at the Oceanic Mesoscale, Annual Review of Marine Science. https://doi.org/10.1146/annurev-marine-010814-015606

McGillicuddy, D.J., Anderson, L. a, Bates, N.R., Bibby, T., Buesseler, K.O., Carlson, C. a, Davis, C.S., Ewart, C., Falkowski, P.G., Goldthwait, S. a, Hansell, D. a, Jenkins, W.J., Johnson, R., Kosnyrev, V.K., Ledwell, J.R., Li, Q.P., Siegel, D. a, Steinberg, D.K., 2007. Eddy/wind interactions stimulate extraordinary mid-ocean plankton blooms. Science (80-. ). 316, 1021–1026. https://doi.org/10.1126/science.1136256

Meyers, G., 1979. On the Annual Rossby Wave in the Tropical North Pacific Ocean. J. Phys. Oceanogr. 9, 663–674.

Mignot, A., Claustre, H., D'Ortenzio, F., Xing, X., Poteau, A., Ras, J., 2011. From the shape of the vertical profile of in vivo fluorescence to Chlorophyll-a concentration. Biogeosciences 8, 2391–2406. https://doi.org/10.5194/bg-8-2391-2011

Mignot, A., Claustre, H., Uitz, J., Poteau, A., Ortenzio, F.D., Xing, X., 2014. Understanding the seasonal dynamics and the deep chlorophyll maximum in oligotrophic environments: A Bio-Argo investigation. AGU. Glob. Biogeochem. cycles 856–876.43

https://doi.org/10.1002/2013GB004781.Received

Mignot, A., Ferrari, R., Claustre, H., 2018. Floats with bio-optical sensors reveal what processes trigger the North Atlantic bloom. Nat. Commun. 9, 1–9. https://doi.org/10.1038/s41467-017-02143-6

Moore, L.R., Goericke, R., Chisholm, S.W., 1995. Comparative physiology of Synechococcus and Prochlorococcus: Influence of light and temperature on growth, pigments, fluorescence and absorptive properties. Mar. Ecol. Prog. Ser. 116, 259–275. https://doi.org/10.3354/meps116259

Muñoz-Anderson, M., Millán-Núñez, R., Hernández-Walls, R., González-Silvera, A., Santamaría-del-Ángel, E., Rojas-Mayoral, E., Galindo-Bect, S., 2015. Fitting vertical chlorophyll profiles in the California Current using two Gaussian curves. Limnol. Oceanogr. Methods 13, 416–424. https://doi.org/10.1002/lom3.10034

Ollitrault, M., Rannou, J.-P., 2013. ANDRO : An Argo-Based Deep Displacement Dataset. J. Atmos 30, 759–788. https://doi.org/10.1175/JTECH-D-12-00073.1

Organelli, E., Claustre, H., Bricaud, A., Schmechtig, C., Poteau, A., Xing, X., Prieur, L., D'Ortenzio, F., Dall'Olmo, G., Vellucci, V., 2016. A novel near-real-time quality-control procedure for radiometric profiles measured by bio-argo floats: Protocols and performances. J. Atmos. Ocean. Technol. 33, 937–951. https://doi.org/10.1175/JTECH-D-15-0193.1

Pennington, J.T., Mahoney, K.L., Kuwahara, V.S., Kolber, D.D., Calienes, R., Chavez, F.P., 2006. Primary production in the eastern tropical Pacific: A review. Prog. Oceanogr. 69, 285–317. https://doi.org/10.1016/j.pocean.2006.03.012

Polito, P.S., Liu, W.T., 2003. Global characterization of Rossby waves at several spectral bands. J. Geophys. Res. 108, 3018. https://doi.org/10.1029/2000JC000607

Polito, P.S., Sato O. T., 2015. Do eddies ride on Rossby waves?. J. Geophys. Res. Oceans. 120, 5417-5435. doi:10.1002/2015JC010737

Poteau, A., Organelli, E., National, I., Xing, X., 2019. Quality control for Biogeochemical-Argo radiometry. https://doi.org/10.13155/62466

Qiu, B., Miao, W., Müller, P., 1997. Propagation and decay of forced and free baroclinic Rossby waves in off-equatorial oceans. J. Phys. Oceanogr. 27, 2405–2417. https://doi.org/10.1175/1520-0485(1997)027<2405:PADOFA>2.0.CO;2

Rasse, R., Dall'Olmo, G., Graff, J., Westberry, T.K., van Dongen-Vogels, V., Behrenfeld, M.J., 2017. Evaluating optical proxies of particulate organic carbon across the surface Atlantic ocean. Front. Mar. Sci. 4, 1–18. https://doi.org/10.3389/fmars.2017.00367

Ravichandran, M., Girishkumar, M.S., Riser, S., 2012. Observed variability of chlorophyll-a using Argo profiling floats in the southeastern Arabian Sea. Deep. Res. Part I Oceanogr. Res. Pap. 65, 15–25. https://doi.org/10.1016/j.dsr.2012.03.003

Rembauville, M., Briggs, N., Ardyna, M., Uitz, J., Catala, P., Penkerc'h, C., Poteau, A., Claustre, H., Blain, S., 2017. Plankton Assemblage Estimated with BGC-Argo Floats in the Southern Ocean: Implications for Seasonal Successions and Particle Export. J. Geophys. Res. Ocean. 122, 8278–8292. https://doi.org/10.1002/2017JC013067

Ripa, P., 2002. Least squares data fitting. Ciencias Mar. 28, 79–105. https://doi.org/doi.org/10.7773/cm.v28i1.204

Roesler, C., Uitz, J., Claustre, H., Boss, E., Xing, X., Organelli, E., Briggs, N., Bricaud, A., Schmechtig, C., Poteau, A., D'Ortenzio, F., Ras, J., Drapeau, S., Haëntjens, N., Barbieux, M., 2017. Recommendations for obtaining unbiased chlorophyll estimates from in situ chlorophyll fluorometers: A global analysis of WET Labs ECO sensors. Limnol. Oceanogr.

Methods 15, 572–585. https://doi.org/10.1002/lom3.10185

Sakamoto, C.M., Karl, D.M., Jannasch, H.W., Bidigare, R.R., Letelier, R.M., Walz, P.M., Ryan, J.P., Polito, P.S., Johnson, K.S., 2004. Influence of Rossby waves on nutrient dynamics and the plankton community structure in the North Pacific subtropical gyre. J. Geophys. Res. C Ocean. 109. https://doi.org/10.1029/2003JC001976

Sathyendranath, S, *et al.,* 2019. An ocean-colour time series for use in climate studies: the experience of the Ocean-Colour Climate Change Initiative (OC-CCI). Sensors. 19, 4285. doi:10.3390/s19194285.

Sauzède, R., Bittig, H.C., Claustre, H., Fommervault, O.P. De, Gattuso, J., Legendre, L., 2017. Estimates of Water-Column Nutrient Concentrations and Carbonate System Parameters in the Global Ocean : A Novel Approach Based on Neural Networks. Front. Mar. Sci. 4, 1–17. https://doi.org/10.3389/fmars.2017.00128

Schmechtig, C., Boss, E.S., Briggs, N.T., Claustre, H., Dall'Olmo, G., Poteau, A., 2019. BGC Argo quality control manual for particles backscattering. https://doi.org/10.13155/60262

Schmechtig, C., Claustre, H., Poteau, A., D'Ortenzio, F., 2018a. Bio-Argo quality control manual for the Chlorophyll-A concentration, Argo Data Management. https://doi.org/10.13155/35385.

Schmechtig, C., Poteau, A., Claustre, H., D'Ortenzio, F., Dall'Olmo, G., Boss, E.S., 2018b. Processing BGC-Argo particle backscattering at the DAC level. https://doi.org/10.13155/57195

Sébastien, L., Josse, J., Husson, F., 2008. FactoMineR: An R Packages for Multivariate Analysis. J. Stat. Softw. 25, 1–18. https://doi.org/10.18637/jss.v025.i01

Siegel, D.A., Jr, D.J.M., Fields, E.A., 1999. Mesoscale eddies , satellite altimetry , and new

production in the Sargasso Sea. J. Geophys. Res. 104, 13,359-13,379. https://doi.org/10.1029/1999JC900051

Stramski, D., Boss, E., Bogucki, D., Voss, K.J., 2004. The role of seawater constituents in light backscattering in the ocean. Prog. Oceanogr. 61, 27–56. https://doi.org/10.1016/j.pocean.2004.07.001

Stramski, D., Reynolds, R.A., Babin, M., Kaczmarek, S., Lewis, M.R., Röttgers, R., Sciandra, A., Stramska, M., Twardowski, M.S., Franz, B.A., Claustre, H., 2008. Relationships between the surface concentration of particulate organic carbon and optical properties in the eastern South Pacific and eastern Atlantic Oceans. Biogeosciences 5, 171–201. https://doi.org/10.5194/bg-5-171-2008

Team, R.C., 2019. R: A language and environment for statistical computing. R Foundation for Statistical Computing.

Uz, B.M., Yoder, J.A., Osychny, V., 2001. Pumping of nutrients to ocean surface waters by the action of propagating planetary waves. Nature 409, 597–600. https://doi.org/10.1038/35054527

Watanabe, W.B., Polito, P.S., da Silveira, I.C.A., 2016. Can a minimalist model of wind forced baroclinic Rossby waves produce reasonable results? Ocean Dyn. 66, 539–548. https://doi.org/10.1007/s10236-016-0935-1

Whitmire, A.L., Letelier, R.M., Villagrán, V., Ulloa, O., 2009. Autonomous observations of in vivo fluorescence and particle backscattering in an oceanic oxygen minimum zone. Opt. Express 17, 21992–22004. https://doi.org/10.3354/meps288035

Wojtasiewicz, B., Trull, T.W., Udaya Bhaskar, T.V.S., Gauns, M., Prakash, S., Ravichandran, M., Shenoy, D.M., Slawinski, D., Hardman-Mountford, N.J., 2018. Autonomous profiling float observations reveal the dynamics of deep biomass distributions in the denitrifying oxygen minimum zone of the Arabian Sea. J. Mar. Syst. 103103. https://doi.org/10.1016/j.jmarsys.2018.07.002

**Figures and tables caption**

**Figure 1.** Representation of main oceanographic features of the ETNP and the trajectory of the BGC-Argo float. Surface chlorophyll-*a* measured by satellite radiometers and geostrophic currents from altimetry were averaged using the 2014-2018 period. The white point represents the deployment location of the float and the black line its trajectory. The position of the North Equatorial Counter Current (NECC), Costa Rica Dome (CRD), Tehuantepec Bowl (TB) and the North Equatorial Current are also indicated. Chlorophylll-*a* data correspond to the version 5 of the ESA-CCI ocean color product (Sathyendranath et al. 2019).

**Figure 2.** Representative vertical profiles gathered by the BGC-Argo float. It is shown examples where the chlorophyll-a presented one (a), two (b) and three maxima (c). It is also shown the vertical distribution of density, dissolved oxygen and particle backscattering (proxy of phytoplankton biomass). Note that chlorophyll-a and particle backscattering maxima coincided in depth.

**Figure 3.** Spatial and temporal evolution of the Sea Level Anomaly (SLA) near the BGC-Argo float position. It is shown the spatial distribution of SLA on May 15th, 2017 (a) and November 16th, 2017 (b). A longitude-time plot at 13.125°N is shown in (c) with two examples of the lines used to estimate the phase speed of Rossby Waves.

**Figure 4.** Temporal evolution of Sea Level Anomaly (SLA; a), chlorophyll-*a* (b) and the particle backscattering coefficient measured at 124° and 700 nm wavelength (bbp700). The position of the 20 °C isotherm (Z20), pycnocline, nitracline and the euphotic zone depth are also shown.

**Figure 5.** Temporal variability of chlorophyll-a (a), the depth of 20C isotherm; Z20 (b) and the Ekman pumping velocity. In (b) **i**s depicted the BGC-float observations of Z20 (dots), their annual harmonic (black line) and the estimated Z20 from a theoretical Rossby wave model (gray thick line).

**Figure 6.** Comparison between the annual cycle from altimetry observations (a-d) and theoretically estimated from a Rossby wave model (e-h). The annual cycle of both datasets was quarterly averaged.

**Figure 7.** Temporal evolution of the Sea Level Anomaly (SLA) and occurrence of eddies on the float position. In (a) it is shown the SLA at the BGC-Argo float position. Each observation of SLA was classified and colored considering if the BGC-Argo float was outside of any Eddy (gray), inside anticyclonic (red points) or within cyclonic eddies (blue). The black line in (a) represents the annual cycle. Three temporal snapshots of the SLA (T1, T2, T3) are represented in the lower panels (b-d). The red and blue contours in the maps delineate spatial boundaries of anticyclonic and cyclonic eddies. We do not screen eddies by lifespan or size.

**Figure 8.** PCA bi-plots for the conditions associated with the oxic-CM (a) and the suboxic-CM (b). Each dot represents an individual observation projected into PCs, which are linear combinations of the original variables: chlorophyll-*a* (chl), particle backscattering coefficient (bbp; proxy of carbon), nitracline depth (z_nit), daily integrated PAR (PAR), dissolved oxygen (DO), and the ratio between bpp700 and chlorophyll-*a* (bbp:chl). Each dot is colored according to the vertical position of the 20 °C isotherm (Z20).

**Figure 9.** Distribution of chlorophyll-a as a function of daily integrated PAR. Chlorophyll-a profiles gathered on "normal" conditions between April-June (AMJ) and October-December (OND) are compared to represent differences along the seasonal cycle (a). The enhancements events observed during November 2017 and December of 2018 are represented in (b) .

**Figure 10.** Time series of daily integrated PAR (a) and the diffuse attenuation coefficient (b) from the BGC-Argo float measurements. Grey points are observations that were smoothed by using a LOESS regression and represented with a solid line in both panels.

**Table 1.** Pearson's correlation coefficients between the Principal Components (PCs) and the original variables describing the conditions at the oxic-CM. Correlations marked with the star symbol (*) were statistically significant at the 95% confidence level.

**Table 2.** Pearson's correlation coefficients between the Principal Components (PCs) and the original variables describing the conditions at the suboxic-CM. Correlations marked with the star symbol (*) were statistically significant at the 95% confidence level.